\documentstyle[epsf,prl,aps,preprint]{revtex}
\newcommand{\mc}{\multicolumn}
\begin{document}

\draft
%\preprint{submitted to Phys.Rev.Lett.}

% -----------------------------------------------------------------------------
\title{Directed Percolation Universality in Asynchronous Evolution of
Spatio-Temporal Intermittency
} 
\author{Juri Rolf,\footnote{Electronic Address: rolf@nbi.dk}
Tomas Bohr,\footnote{Electronic Address: tbohr@nbi.dk} and
Mogens H. Jensen\footnote{Electronic Address: mhjensen@nbi.dk}}
\address{Niels Bohr Institute 
and Center for Chaos and Turbulence Studies, 
Blegdamsvej 17, DK-2100 {\O}, Denmark }
\date{June 23, 1997}
\maketitle
% -----------------------------------------------------------------------------
\begin{abstract}
We present strong evidence that a coupled-map-lattice model 
for spatio-temporal intermittency
belongs to the universality class of directed percolation when the 
updating rules are {\em asynchronous}, i.e. when only one randomly chosen 
site is evolved
at each time step. In contrast, when the system is subjected to parallel 
updating, available numerical evidence suggest that it does not belong to 
this universality class and that it is not even universal. We argue that in 
the absence of periodic external forcing, the asynchronous rule is the more 
physical.
\end{abstract}
The onset of spatio-temporal intermittency is a common phenomenon 
of many extended
systems ranging from models based on 
coupled-map-lattices \cite{Chate1,Chate3,Chate2}
to various experiments in convection \cite{Ciliberto,Berge} and in 
the ``printers instability'' \cite{Rabaud,Michalland}. A
particular elegant coupled-map-lattice (CML) showing 
spatio-temporal intermittency
was introduced some years ago by Chat\'e and Manneville
\cite{Chate1,Chate3}. This CML employs individual maps that can be in two
very different states: either in a chaotic (or ``turbulent'') state
or in a ``laminar'' state. For a single map
the laminar state is ``absorbing'': once the motion is in the laminar
state it cannot escape. For the coupled system, one observes
interesting dynamical patterns of turbulent patches penetrating into 
a laminar state, and because of the strong fluctuations, this has been
called \cite{Chate1} spatio-temporal intermittency (STI). 
Once the system is in a pure
laminar state, it cannot escape: this is an absorbing state
of the full spatially coupled system. These properties of the 
STI led Pomeau \cite{Pomeau} to conjecture that the 
critical properties
at the onset of STI should be governed by the exponents of
directed percolation. The turbulent spots of the dynamics in a space-time
plot percolate through the system in a manner very similar
to the connected bonds in directed percolation (DP), which
also has an absorbing state.
Subsequent extensive numerical studies and scaling arguments
\cite{Chate3,Chate2,Houlrik1} did not show agreement with this conjecture.
On the contrary, it was found that the
generic critical properties were not in the universality 
class of directed percolation. In fact, since the critical properties vary 
with the parameters, they are not even universal.
Nevertheless it was argued in \cite{Grassberger} that the
apparent non-universality is due to travelling solitary
excitations with long life times,
and that one should in principle recover the DP-behaviour only
for extremely long time scales.

The standard time evolution of a coupled-map-lattice
is by synchronous (or parallel) updating, in which all individual
maps of the lattice are iterated forward simultaneously in a
completely deterministic way \cite{Kaneko}. The connection between such 
coupled map lattices and physical systems, described by partial differential 
equations is, however, not straight forward. In particular, the synchronous 
update can only be motivated when the system dynamics 
is driven by a global periodic external force (or ``clock'').
Recently it was observed that the critical properties 
of a standard CML, using coupled logistic maps, actually depend 
on whether synchronous or asynchronous updating is applied \cite{Marcq}. 
Here, asynchronous means that in each step a random
site on the CML is chosen and is iterated forward. 

This result forces us to reconsider the evidence for non-universal  
behaviour for STI, which is all based on the synchronous updating rule.
In this letter, we thus 
apply the asynchronous updating to the coupled-map-lattice
of spatio-temporal intermittency discussed above. We find that all
critical exponents, independent of the choice of parameters,
fall into the universality class of directed percolation. All
critical characteristics of DP, such as hyper-scaling, seem to be fulfilled,
leading to independent controls of the values of the critical exponents.
%We believe that this resolves a long-standing puzzle of the relation between 
%directed percolation and the dynamics of STI.

The original dynamics of the coupled-map-lattice of Chat\'e and Manneville
\cite{Chate1} for a lattice with one space- and one time-direction
is written in terms of a field $u_i(t)$ at site $i$ and time $t$ as
%The original coupled-map-lattice of Chat\'e and Manneville 1d$d$
%\cite{Chate1} 
%is written in terms of a field $u_i(t)$ at site $i$ and time $t$
\begin{equation}
\label{eq1}
u_i(t+1) = f(u_i(t)) +
 {\epsilon \over 2} \Delta_fu_i(t)\, ,
\end{equation}
where $\Delta_fu_i(t)=f(u_{i-1}(t))-2f(u_i(t))+f(u_{i+1}(t))$.
The parameter $\epsilon$ measures the strength of the coupling from site $i$
to its two neighbours. 
The dynamics (\ref{eq1}) is parallel or synchronous:
all sites are updated at the same time.

The local map $f$ is of the form:
\begin{equation}
\label{eq2}
   f(x) = \cases{rx,     & if $x\in\lbrack 0,1/2\rbrack$; \cr
                 r(1-x), & if $x\in\lbrack 1/2,1\rbrack$; \cr
                 x,      & if $x\in\rbrack 1,r/2\rbrack$. \cr}
\end{equation}
The chaotic motion of $f$ for $x \le 1$ is governed by a standard tent
map of slope $r$. However, when $r$ exceeds the value two, the 
trajectory may escape to a ``laminar'' state with $x > 1$, 
and this state is marginally stable, because the slope in the line of fixed 
points is one. As mentioned above,
the laminar state is absorbing, i.e. the trajectory cannot be pulled back
into the chaotic state. This is no longer the case,  when the maps are 
coupled, since the interactions with its neighbours may
pull a laminar site back into chaotic motion, thus causing the
interesting interplay between laminar and turbulent regions.
%The temporal dynamics of (\ref{eq1}) is written as synchronous or parallel:
%all sites are updated at the same time.

In this work we consider the same CML
under asynchronous updating: at time $t$ a random site $i_r$ is selected
and is altered according to (\ref{eq1}) while all the other variables 
keep their values:
\begin{eqnarray}
\label{eq3}
u_{i_r}(t+1/L) &=& f(u_{i_r}(t)) +
 {\epsilon \over 2} \Delta_fu_{i_r}(t)\, ,\\
u_i(t+1/L) &=& u_i(t)~~~~{\rm for}~~~~ i \neq i_r\, ,\nonumber
\end{eqnarray} 
where $L$ is the size of the system. Note that with this choice of time step
on average each site is 
updated once in one unit of time.

The introduction of chaotic iterated maps in extended dynamical
systems - and especially simple one-dimensional, non-invertible maps -
can only be motivated in a very heuristic way \cite{Chate1}. The closest
physical parallel is probably a collection of weakly coupled subunits,
each able to perform chaotic dynamics.
In order to derive a discrete, local map, one applies
a Poincar\'e section.
Note that the Poincar\'e map owes its simplicity to
the fact that it is obtained by restricting the dynamics to a surface
through the local phase space and is therefore not in general
equivalent to moving the system forward through a fixed time interval.
In the absence of an external, periodic forcing, the time interval
between consecutive crossings of the section will generically 
be variable and thus
vary from unit to unit in space. To find the state
of the entire system at {\it fixed} time intervals, this variation
has to be taken into account. This can, in a
rough way, be accomplished by using the asynchronous update in which
the units experience slightly different update times. Of course we are
thereby replacing the complicated, deterministic update rule by a
random one and this is obviously a crude approximation. But we believe
that this is often closer to physical reality than the synchronous update.
As an example, synchronous updating can lead
to complicated, spatial structures, in cellular automata \cite{May},
which disappear under asynchronous
updating \cite{Huberman}, and should therefore be considered unphysical
in the absence an external clock.

\begin{figure}[htbp]
  \begin{center}
    \epsfxsize=12cm
    \epsffile{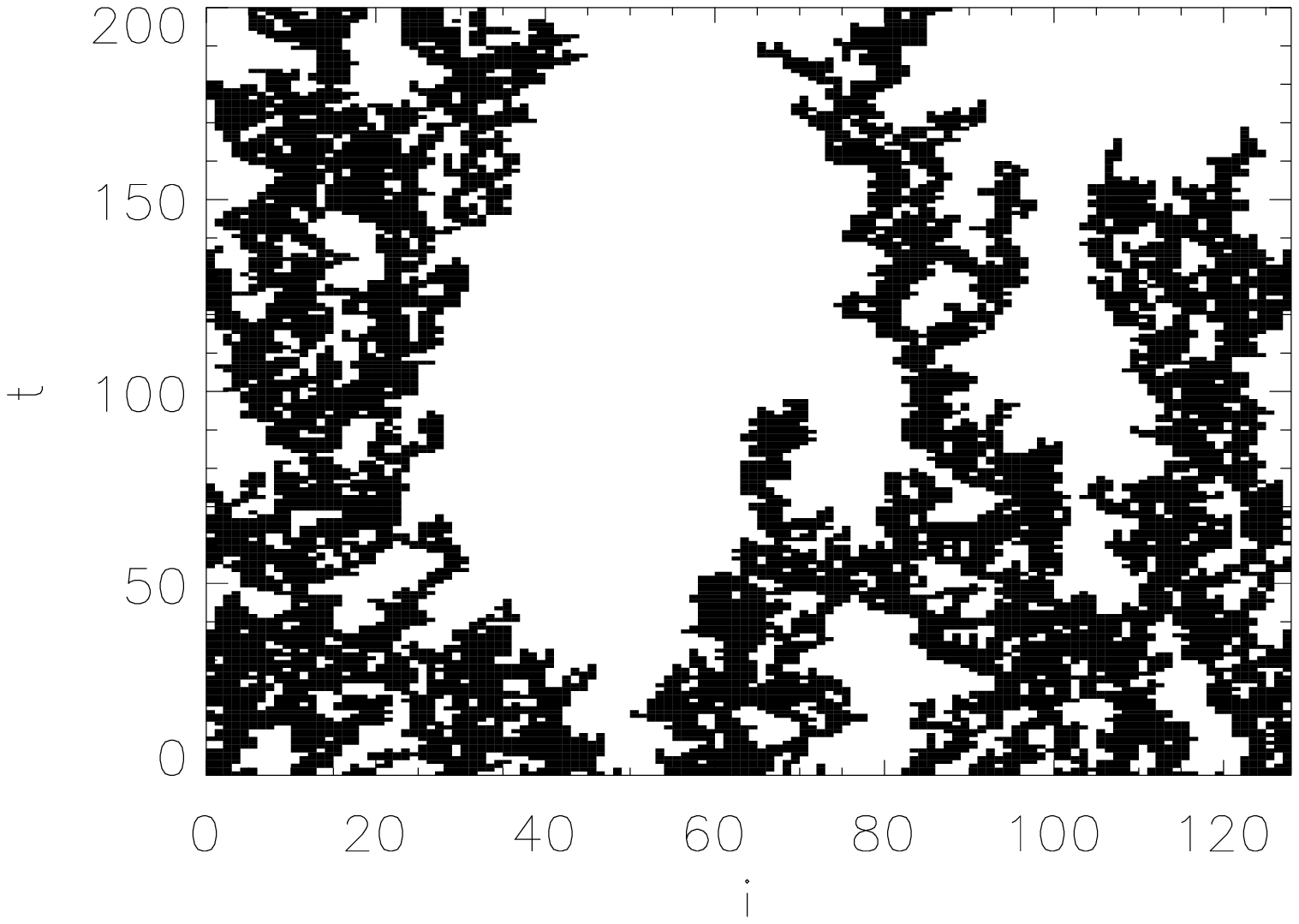}
  \end{center}
  \caption[x]{Time evolution of the asynchronous CML (\ref{eq3}) 
    with $r=3.0$,  $\epsilon=0.58$ and system size $L=128$.
    The turbulent sites with $u \leq 1$ are black while
    the laminar sites with $u > 1$ are white.}
\label{fig:1}
\end{figure}

Fig. \ref{fig:1} shows a pattern generated by the asynchronous update
introduced above. As only one site is
updated at a time, a new horizontal line in the time axis is added
only after $L=128$ time steps (i.e. $t \rightarrow t+1$). 
We observe that the turbulent sites (black
in Fig. \ref{fig:1}) percolate through the system, sometimes ending in a
dangling bond.
%%The boundaries of these penetrating patterns are less
%%regular than for the synchronous case \cite{Chate1}. In fact this type
%%of pattern is closer to the experimentally observed 
%%STI patterns than the patterns generated with synchronous updating.

\begin{figure}
  \begin{center}
    \epsfxsize=12cm
    \epsffile{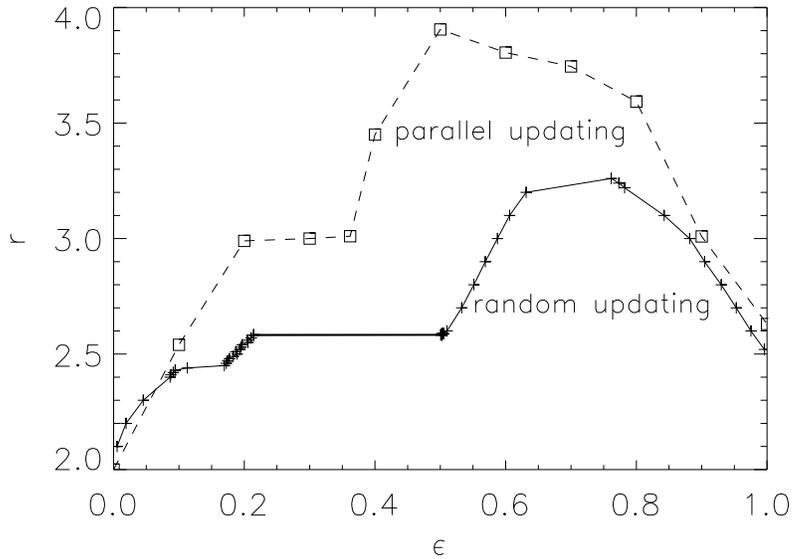}
  \end{center}
  \caption[x]{Phase diagrams of one-dimensional CML
    for asynchronous updating (crosses) and synchronous updating (boxes).}
  \label{fig:2}
\end{figure}

To estimate the critical exponents for the randomly updated CML 
(\ref{eq3}), we
follow the finite size scaling methods of Houlrik et al. 
\cite{Houlrik1,Houlrik2,Houlrik3}.
First of all we have to locate the critical line
in the parameter plane $(\epsilon , r )$. This is done
by measuring the absorption time $\tau (r,\epsilon,L)$,
i.e. the 
time it takes the system starting from a random initial state to reach
the absorbing state - averaged over an ensemble of initial conditions.  
At the critical point $\epsilon=\epsilon_c(r)$, this time diverges like
\begin{equation}
\label{eq4}
\tau (\epsilon_c , L) \sim L^z
\end{equation}
where the usual dynamical exponent $z = \nu_{\parallel} /
\nu_{\perp}$ has been introduced. Fig.~\ref{fig:2} shows the phase diagram 
with
the critical line and contrasts it with the critical line for the
synchronously updated system (\ref{eq1}) taken from \cite{Houlrik1}.
We consider three different
values of $r$ and the corresponding values of $\epsilon_c$ and $z$
are found in Table \ref{table1}. 

\begin{figure}[htbp]
  \begin{center}
    \epsfxsize=12cm
    \epsffile{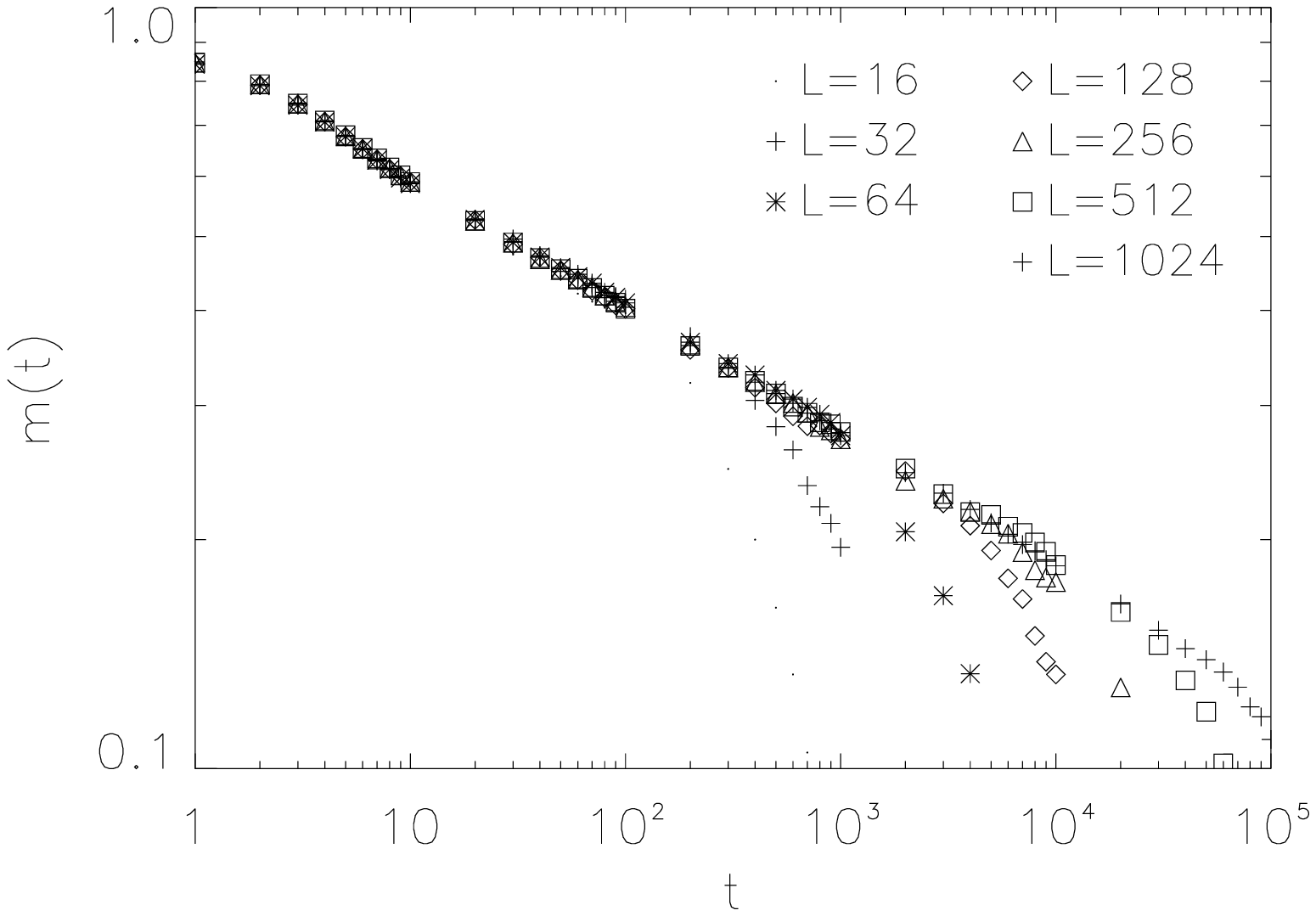}
  \end{center}
  \caption[x]{Order parameter $m(t)$ versus time $t$ for system sizes
    $L$ between $2^4$ and $2^{10}$ at $r=2.2$. As expected the data
    falls on one line as long as $t$ is much smaller than the escape time.}
\label{fig:3}
\end{figure}

The order parameter $m(\epsilon,L,t)$ is
defined as the fraction of turbulent sites in the lattice,
again averaged over many different initial states \cite{Houlrik1}.
The order parameter scales in the usual way when approaching
the critical line from above
\begin{equation}
\label{eq5}
m \sim (\epsilon - \epsilon_c)^{\beta}~~~~{\rm for}~~~~
\epsilon \to \epsilon_c^+
\end{equation}
Using this relation we have estimated the $\beta$-exponent directly
and the results are found in Table \ref{table1}.
Applying finite-size scaling arguments we find the following
scaling form at the critical point
\begin{equation}
\label{eq6}
m(\epsilon_c,t,L) \sim L^{-\beta/\nu_{\perp}} g(t / L^z).
\end{equation}
The function $g(t / L^z)$ has the following properties: At times
much smaller than the absorption time, one expects a power-law decay
in time due to critical correlations and the $L$-dependent 
pre-factor in (\ref{eq6}) must drop out. For $t \ll L^z$, we therefore have
\begin{equation}
\label{eq7}
m(\epsilon_c,t,L) \sim t^{-\beta/\nu_{\parallel}}
\end{equation}
For times much larger that the absorption time, we may assume
uncorrelated decay of the order parameter, and therefore
the function $g$, will decay exponentially.
Fig.~\ref{fig:3} shows a plot of $m(\epsilon_c,t,L)$ versus $t$ at $r=2.2$. The
curves for 7 different system sizes in the interval $L=2^4 ,...., 2^{10}$
fall very accurately on the same line in the double logarithmic plot
allowing
determination of $\beta/\nu_{\parallel}$ as listed in Table \ref{table1}.
As we see in this table all the critical exponents are, within
the error bars, consistent with the values for directed percolation, listed
at the bottom line in Table \ref{table1}. 

\begin{figure}[htbp]
  \begin{center}
    \epsfxsize=12cm
    \epsffile{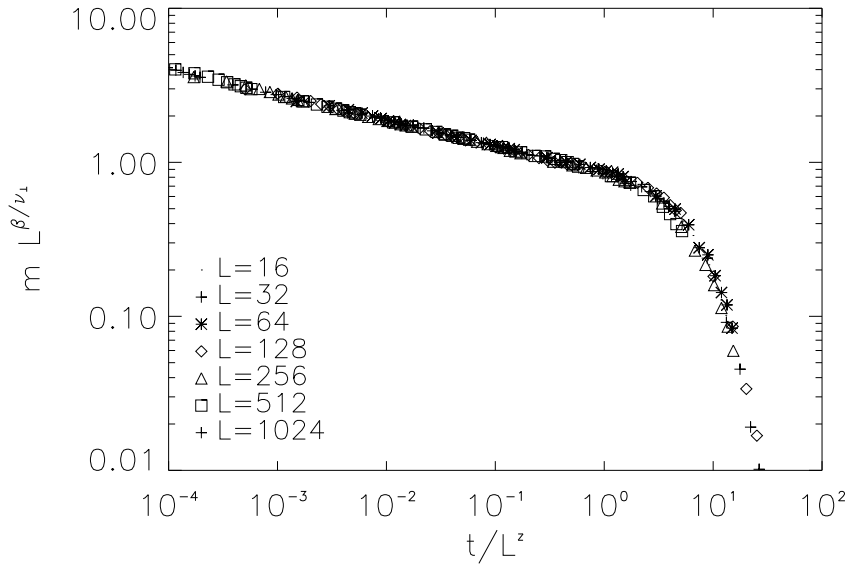}
  \end{center}
  \caption[x]{Rescaling of the order parameter at $r=2.2$ according to
    (\ref{eq6}) to obtain estimates for
    $\beta/\nu_{\perp}$ and for $z$. Data for
    for system sizes $L$ between $2^4$ and $2^{10}$ collapses
    on one curve.}
  \label{fig:4}
\end{figure}

In order to obtain independent checks on the values of the critical 
exponents we have performed a rescaling analysis using (\ref{eq6}).
The rescaled
curves are shown in Fig.~\ref{fig:4}. For systems sizes 
in the interval $L=2^4 ,...., 2^{10}$ the rescaled
curves collapse very accurately 
to a single curve.
The corresponding values of 
$\beta / \nu_{\perp}$ and $z$ are shown in Table \ref{table2}- again
consistent with DP.

\begin{figure}[htbp]
  \begin{center}
    \epsfxsize=12cm
    \epsffile{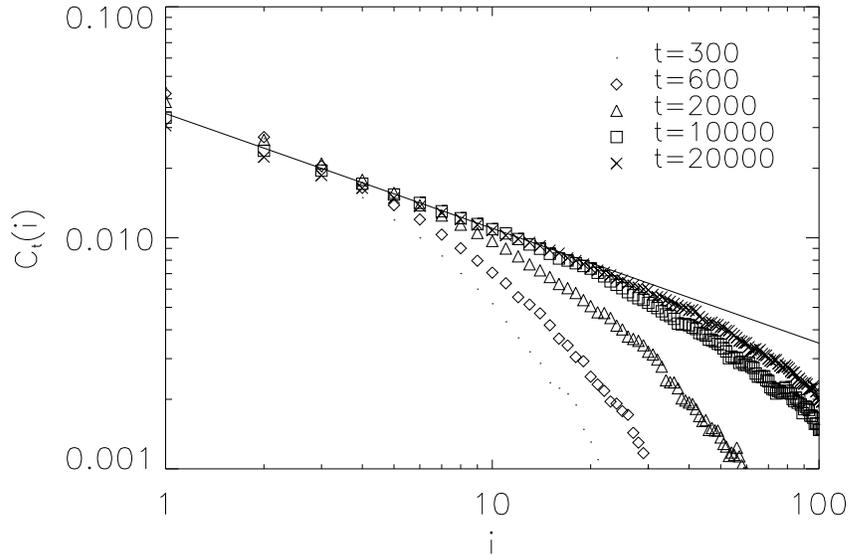}
  \end{center}
  \caption[x]{Spatial correlation function $C_i(t)$ for
    various times $t$ at $r=2.2$.
    $C_i(t)$ approaches a straight line with slope $1-\eta$ for large times,
    indicating an algebraic decay.}
  \label{fig:5}
\end{figure}

To extract further critical exponents and to check 
hyper-scaling we have looked at the spatial
correlations. Even though any dynamics of this CML will end up in
the absorbing (laminar) state, it is possible to find non-trivial
spatial correlations in a long lasting quasi stationary situation
\cite{Houlrik2}. The correlations can be obtained from the
pair correlation function
\begin{equation}
\label{eq8}
   C_j(t) = {1\over L}\sum_{i=1}^L \langle u_i(t) u_{i+j}(t) \rangle
-   \langle u(t) \rangle ^2\, ,
\end{equation}
where the brackets denote the average over different initial conditions. 
If the coupling between the sites is weak, i.e. if $\epsilon$ is
small, one 
might expect the spatial correlations to fall off exponentially
with distance. At criticality, on the other hand, one expects
an algebraic decay of correlations \cite{Houlrik2}
\begin{equation}
   C_j(t) = j^{1-\eta} \, \psi(j/\xi(t))~,
\end{equation}
where $\eta$ is the associated critical exponent, and correlations
are induced over a length scale $\xi(t)\sim t^{1/z}$ as the CML
relaxes towards a steady state.
The (equal time) correlation function for various times
(at $r=2.2$) is shown in Fig. \ref{fig:5}, 
indicating an algebraic decay in space after long 
time. The
corresponding value of $\eta$ from this direct measurements 
is shown in Table \ref{table1} together with values obtained at other 
$r$-values.
Again the agreement with DP is confirmed. 

\begin{figure}[htbp]
  \begin{center}
    \epsfxsize=12cm
    \epsffile{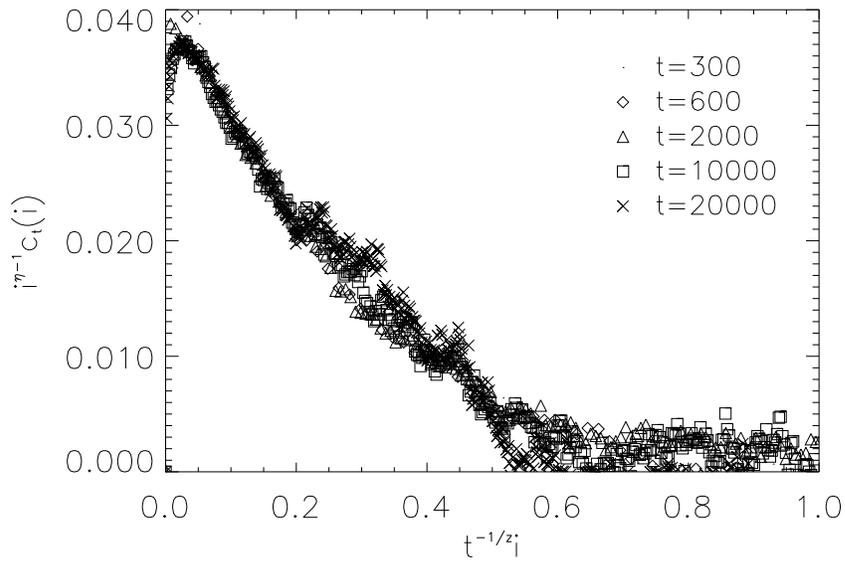}
  \end{center}
\caption[x]{Rescaled spatial correlation function at $r=2.2$. The data for
  various times collapses on one curve if the exponents $z$ and $\eta$
  take the values in table \ref{table2}.}
\label{fig:6}
\end{figure}

For the spatial correlations, 
a rescaling analysis can also be performed by plotting
$j/t^{1/z}$ versus $j^{\eta-1}C_j(t)$. Fig. \ref{fig:6} shows the corresponding
rescaled plot and this technique allows another independent
determination of $z$ and $\eta$, the values of which are shown
in Table \ref{table2}.

The fourth column of Table \ref{table2} contains the values of 
$\eta$ obtained from the
hyper-scaling relation
\begin{equation}
\label{hyper}
2\beta/\nu = d-2+\eta~,
\end{equation}
giving a third way of estimating the exponent $\eta$. All three ways of 
finding $\eta$ give results that, within error, are in
agreement with the directed percolation value 1.51. 

We thus conclude that our numerics gives very strong evidence for the
fact that spatio-temporal intermittency in the form of asynchronous
coupled maps falls in the universality class of directed percolation.
The reason why synchronously updated maps do not behave in a universal 
way must thus be sought in the complicated correlations built up
by the strong constraint of exactly simultaneous, completely deterministic
updating, which,
in most applications will not be fulfilled. Thus, experiments 
showing spatio-temporal intermittency should be describable by 
directed percolation near the transition to turbulence - at least 
insofar as they do not involve global periodic forcing. One might speculate
on the possibility of observing the non-universal ``synchronous" behaviour
in periodically forced systems, as say the printers instability driven by
a cylinder of non-circular cross section.

\acknowledgements
We would like to thank Bernardo Huberman and Benny Lautrup for helpful 
discussions.
J.R. gratefully acknowledges financial support by the
Studienstiftung des deutschen Volkes.

% --------------------------------------------------------------------
% BIBLIOGRAPHY
% --------------------------------------------------------------------

% --------------------------------------------------------------------
% FIGURE CAPTIONS
% --------------------------------------------------------------------
\epsfclipoff
\begin{table}[htbp]
    \caption{Direct measurements of the critical exponents for system 
(\ref{eq3}).
The critical values $\epsilon_c$ and the exponent $z$ are found
simultaneously by approaching the value of $\epsilon$ where the
scaling (\ref{eq4}) is best. The estimation of the other exponents are
described in the text. The bottom line are the directed percolation 
exponents.}
    \label{table1}
    \begin{tabular}{dddddd}
      \mc{1}{c}{$r$} &
      \mc{1}{c}{$\epsilon_c$} &
      \mc{1}{c}{$z$} &
      \mc{1}{c}{$\beta$} &
      \mc{1}{c}{$\frac{\beta}{\nu_{\parallel}}$} &
      \mc{1}{c}{$\eta$} \\ \hline
      2.2 & 0.0195(2)   & 1.58(2) & 0.28(1) & 0.16(1) & 1.51(2) \\
      2.6 & 0.5096(2)   & 1.59(2) & 0.28(1) & 0.16(1) & 1.49(2) \\
      3.0 & 0.5870(3)   & 1.60(3) & 0.28(1) & 0.15(2) & 1.50(2) \\
      \mc{1}{c}{DP}& & 1.57    & 0.28    & 0.16    & 1.51    \\
    \end{tabular}
\end{table}
\begin{table}[htbp]
    \caption{The exponents and relations obtained from rescaling analysis 
using finite
size scaling. 
The fourth column is an estimate of $\eta$ using the hyper-scaling relation
(\ref{hyper}).}
    \label{table2}
    \begin{tabular}{dddddd}
      \mc{1}{c}{$r$} &
      \mc{1}{c}{$\frac{\beta}{\nu_{\perp}}$} &
      \mc{1}{c}{$z$} &
      \mc{1}{c}{$\eta = 2 \frac{\beta}{\nu_{\perp}} + 1$} &
      \mc{1}{c}{$z$} &
      \mc{1}{c}{$\eta$} \\ \hline
      2.2 & 0.26(1) & 1.57(1) & 1.52(2) & 1.56(2) & 1.53(2)\\
      2.6 & 0.26(1) & 1.57(1) & 1.52(2) & 1.58(2) & 1.49(2)\\
      3.0 & 0.25(2) & 1.58(1) & 1.50(3) & 1.58(2) & 1.53(2)\\
      DP &  0.26    & 1.57    & 1.51    & 1.57    & 1.51   \\
    \end{tabular}
\end{table}
% --------------------------------------------------------------------
\end{document}